\documentclass[twocolumn,prb,aps,letterpaper]{revtex4}
\usepackage{amsmath,amssymb,bm,revsymb,graphicx,verbatim}

\begin{document}
\newcommand{\ket}[1]{\ensuremath{
				|#1\rangle}
			}
\newcommand{\bra}[1]{\ensuremath{
				\langle#1|}
			}
\newcommand{\Threej}[6]{\ensuremath{
				\begin{pmatrix}
					#1 & #2 & #3\\
					#4 & #5 & #6
				\end{pmatrix}}
			}

\newcommand{\threej}[6]{\ensuremath{
				\bigl(
				\begin{smallmatrix}
					#1 & #2 & #3\\
					#4 & #5 & #6
				\end{smallmatrix}
				\bigr)}
			}

\newcommand{\Sixj}[6]{\ensuremath{
				\begin{Bmatrix}
					#1 & #2 & #3\\
					#4 & #5 & #6
				\end{Bmatrix}}
			}

\newcommand{\sixj}[6]{\ensuremath{
				\bigl\{
				\begin{smallmatrix}
					#1 & #2 & #3\\
					#4 & #5 & #6
				\end{smallmatrix}
				\bigr\}}
			}

\newcommand{\unit}[1]{\ensuremath{
				\mathbf{{e}}_{#1}
				}
			}

\newcommand{\Unit}[1]{\ensuremath{
				\mathbf{{e}}^{#1}
				}
			}

\begin{abstract}
The Rabi frequency (coupling strength) of an electric-dipole transition is an
important experimental parameter in laser-cooling and other atomic physics
experiments.  Though the relationship between Rabi frequency and atomic
wavefunctions and/or atomic lifetimes is discussed in many references, there is
a need for a concise, self-contained, accessible introduction to such
calculations suitable for use by the typical student of laser cooling
(experimental or theoretical). In this paper, I outline calculations of the Rabi
frequencies for atoms with sub-structure due to orbital, spin and nuclear
angular momentum. I emphasize the \emph{physical meaning} of the calculations.
\end{abstract}

\pacs{}
\draft
\title{Angular Momentum Coupling and Rabi Frequencies for Simple Atomic
Transitions}
\author{B.E.~King}
\address{Dept. Physics \& astronomy, McMaster University, Hamilton ON}
\date{\today}
\maketitle

\section{Introduction}
In designing or implementing many modern atomic physics experiments (e.g. laser
cooling, magneto-optical trapping, dipole trapping, optical pumping, etc.) it
is important to be able to calculate the coupling strength or Rabi frequency of
a laser-driven transition between two atomic states.  However, a first attempt
to do this can be a frustrating experience.

Many of the older books on atomic spectroscopy were written at a time when
coherent excitation was not possible; for this reason these works often focus on
multi-line excitation or on spontaneous emission from many, thermally excited
levels.  In addition to the pedagogical barrier this may present, one may also
have to surmount the obstacles of older notations for quantum states or the use
of CGS units. Though translating to more modern usage is straightforward in
principle, in practice it can be a confusing endeavour.  More modern textbooks
which treat laser-atom interactions at an introductory level are either aimed at
laser dynamics, treat the atoms as two-level systems, or only sketch out the
calculations.\footnote{Metcalf's book on Laser Cooling.\cite{CoolingNeutrals}
does an excellent job outlining the physics of laser coupling in multi-level
atoms, but - no doubt for brevity - skips the details.}

In fact, there are no two-level atoms, and so practical calculations of Rabi
frequencies are more involved.  However, the fact that the atomic states
have well-defined symmetries - as embodied by the \emph{Wigner-Eckart theorem} -
allows for considerable simplification in the calculations. In this paper, I
attempt to provide a pedagogical overview of Rabi-frequency
calculations for multi-level atoms.  Wherever possible, I try to provide
physical pictures corresponding to the math.  Since the majority of
laser-cooling experiments are performed on hydrogen-like atoms (e.g. $Li$, $Na$,
$K$, $Rb$, $Cs$, $Be^+$, $Mg^+$, $Cd^+$,...), I treat only
single-electron excitation of atoms with such a configuration.  Likewise, since
the laser-atom interactions are typically electric-dipole transtions, I only
calculate Rabi frequencies for such transitions.  Though the generalization of
the calculations to magnetic dipole, electric quadrupole transitions, etc., is
straightforward, the reader is referred to the literature for such calculations.

In Sec. \ref{dipole}, I review the interaction between a
linearly polarized laser field and a two-level atom in the
\emph{rotating-wave} and \emph{dipole approximations}, as parametrized by the
Rabi frequency.  Next (Sec. \ref{degeneracy}), I outline the repurcussions of
degeneracy.  After a brief overview of rotational symmetry, angular momentum,
and their quantum implications in Sec. \ref{symmetry}, I treat the combination
of angular momenta in Sec. \ref{coupling}.  In Sec. \ref{Wigner_Eckart},
I explain how the \emph{Wigner-Eckart theorem} can simplify calculations for
transitions between states of well-defined angular momentum, if the transitions
may be represented in terms of operators with well-defined rotational symmetry
(i.e. as \emph{tensor operators}).

Using the Wigner-Eckart theorem, I relate the Rabi frequencies of transitions
between various angular-momentum sublevels to the excited-state lifetime.  I
first treat states with well-defined total angular momentum $\mathbf{\hat{J}}$
(Sec. \ref{J}), before breaking down the explicit dependence upon orbital
angular momentum $\mathbf{\hat{L}}$ and radial overlap integrals
$\mathcal{R}_{n'l'}^{nl}$ (Sec. \ref{ls}). Finally, I discuss the case of atoms
with nuclear spin and hyperfine structure in Sec. \ref{HFS}.  The main results
of this paper are Eq. (\ref{expt_Rabi}) (which expresses the Rabi frequency in
terms of the laser beam's electric field, intensity, and the power/beam waist of
a Gaussian beam) and Eq. (\ref{actual_Rabi}), Eq. (\ref{decoupled_Rabi}), and
Eq. (\ref{hyperfine_Rabi}), which relate the Rabi frequency to the atom's
lifetime and the electric field of the laser in various angular momentum
coupling schemes.

This present work attempts to provide the bare minimum of material necessary
for the reader to understand and calculate the Rabi frequency for simple
cases.  In an attempt to save the reader an exhaustive literature search,
I have, wherever possible, drawn mathematical results from a single
source - Messiah's canonical text on quantum mechanics.\cite{Messiah}  Metcalf
and van der Straten's book places this calculation in the context of laser
cooling and trapping of neutral atoms\cite{CoolingNeutrals} For further
background, Cowan's\cite{Cowan_book} or Weissbluth's\cite{Weissbluth} books
provide excellent reading. Finally, Suhonen\cite{Suhonen} gives a succinct
review of angular momentum and irreducible tensor operators, while
Silver\cite{Silver_tensors} provides further dicussion of rotational symmetry
and tensors.

\section{The Rabi Frequency}
\label{Rabi_sec}

\subsection{The Dipole Interaction with two-level atoms}
\label{dipole}

Let us begin by considering the case of a (fictional!) atom
with only two levels:  ground state $|g\rangle$ and excited state
$|e\rangle$.  Let the energies of these levels be $E_g$ and
$E_e$, respectively, and let $\omega_0=(E_e-E_g)/\hbar$.  In general, the state
of the atom may be written as $|\Psi\rangle=c_g(t)|g\rangle+c_e(t)|e\rangle$,
where $|c_g(t)|^2+|c_e(t)|^2=1$.

Suppose that one applies to this atom a resonant, linearly-polarized laser field
of the form
${\mathbf{E}}({\mathbf{r}},\;t)={\bm{\epsilon}}\;E_0\;\cos({\mathbf{k}
} \cdot{\mathbf{r}}-\omega_L t)$.  Here $E_0$ is the electric-field amplitude,
$\mathbf{k}$ is the wavevector, $\omega_L$ is the angular frequency of the
laser, which we take to be equal to $\omega_0$, and $\bm{\epsilon}$ is a unit
vector in the direction of polarization ($\bm{\epsilon}\perp\mathbf{k}$).  The
basis vectors for the atomic Hilbert space are themselves evolving with time
dependence $e^{-iE_nt/\hbar}$, so calculations are easier if we rewrite the
laser field in complex form: 
${\mathbf{E}}({\mathbf{r}},\;t)=\frac{1}{2}{\bm{\epsilon}}\;E_0\;\left[e^{i({
\mathbf { k }} \cdot{\mathbf{r}}-\omega_L t)}+e^{-i({
\mathbf { k }} \cdot{\mathbf{r}}-\omega_L t)}\right].$  Since we take
$\omega_0$, $\omega_L$ to be positive,
$(\omega_L+\omega_0)\gg(\omega_L-\omega_0)$ and one often makes the
\emph{rotating wave approximation} of dropping the second
exponential.\footnote{This approximation is responsible for the factors of
$\frac{1}{2}$ appearing below in the state vectors' time evolution.}

Under the assumptions that the laser interaction is weak compared to atomic
effects and that the size of the atom is much less than the wavelength of light,
we may make the electric-dipole approximation:\cite{cohent, Woodgate}  the
interaction Hamiltonian is given by
$\hat{V}_I=-\;\bm{\hat{\mu}\cdot}\mathbf{E}$. Here
$\bm{\hat{\mu}}=-e\bm{\hat{r}}$ is the dipole operator for the atom and $e$
is magnitude of the charge on the electron. The result of the interaction is
that $|g\rangle$ and $|e\rangle$ become coupled.

Suppose that the atom is initially in the state $|g\rangle$.  If we neglect
spontaneous emission from $|e\rangle$, then under the rotating-wave
approximation and in the Schr\"{o}dinger representation, the time dependence of
the system is given by

\begin{subequations}
\label{Rabi_flopping}
\begin{align}
	c_g(t) &= e^{-iE_g t/\hbar}\;\cos\left(\frac{\Omega t}{2}\right) \\
	c_e(t) &= e^{-iE_e t/\hbar}\;\sin\left(\frac{\Omega t}{2}\right).
\end{align}
\end{subequations}

\noindent (The exponential terms show the time evolution due to the bare atomic
Hamiltonian.)  Here, I have defined the \emph{Rabi frequency} of the transition
to be

\begin{equation}
	\Omega:=-\frac{\langle
		e|\bm{\hat{\mu}\cdot\epsilon}\,E_0|g\rangle}{\hbar}.
\end{equation}

\noindent 
The Rabi frequency measures the strength of the coupling between the atomic
states and the applied electromagnetic field.  Practically speaking, one
doesn't directly measure the electric field amplitude, but rather the
peak intensity $I=\frac{1}{2}\epsilon_0 c E_0^2$ of the laser beam or, more
typically, the total power $P$ and beam waist $w_0$ of a Gaussian laser beam
($I=\frac{2P}{\pi w_0^2}$). 
Thus:

\begin{subequations}
\label{expt_Rabi}
\begin{align}
	\Omega  & = \frac{eE_0}{\hbar}\langle
		e|\bm{\hat{r}\cdot\epsilon}|g\rangle \\
		& = \sqrt{\frac{e^2 2I}{\epsilon_0\hbar^2 c}}\langle
		e|\bm{\hat{r}\cdot\epsilon}|g\rangle \\
		& = \sqrt{\frac{4e^2 P}{\epsilon_0\pi\hbar^2 cw_0^2}}\langle
		e|\bm{\hat{r}\cdot\epsilon}|g\rangle		
\end{align}
\end{subequations}

Eq. (\ref{Rabi_flopping}) indicates that the state vector of the system
oscillates coherently between $|g\rangle$ and $|e\rangle$ with frequency
$\Omega/2$ - a behaviour which is called \emph{Rabi flopping}.  On the
other hand, the populations oscillate as
$|c_g|^2=\cos^2\left(\frac{\Omega
t}{2}\right)=\frac{1}{2}\left[1+\cos\left(\Omega t\right)\right]$ and
$|c_e|^2=\sin^2\left(\frac{\Omega
t}{2}\right)=\frac{1}{2}\left[1-\cos\left(\Omega t\right)\right]$.  So according
to Eq. (\ref{Rabi_flopping}), the Rabi frequency is the frequency at which the
\emph{populations} oscillate.\footnote{There isn't universal agreement as to the
definition of the Rabi frequency.  While many sources\cite{CoolingNeutrals,
Farell_MacGillivray, Meystre_Sargent, Haken_Wolf, atom-phot-int, Loudon,
Basdevant_DalibardQM, Zubairy, Demtroder, Allen_Eberly, Foot, Walls_Milburn}
use the definition of Eq. (\ref{expt_Rabi}), others \cite{Verdeyen, Siegman,
King_thesis} define the Rabi frequency to be \emph{one-half} of $\Omega$. 
However, the definition in question can always be determined by comparing the
Rabi-flopping equations with Eq. (\ref{Rabi_flopping}).}  A pulse for which
$\Omega t=\pi$ is called a ``pi pulse'' - it results in complete population
transfer to the excited state.  Similarly a pulse for which $\Omega t=\pi/2$ is
called a ``pi-by-two pulse'' and results in an equal superposition of ground and
excited states.  Note that, though a ``two-pi pulse'' ($\Omega t=2\pi$) returns
the \emph{population} to the ground state, it takes $\Omega t=4\pi$ to return
the \emph{state vector} to its initial value.  (This is analagous the the
requirement of a $4\pi$ rotation to return a spin-$\frac{1}{2}$ particle to its
initial state.)

A perturbation-theory approach to the problem (again, in the rotating-wave
approximation) predicts that, for short times, before the population of the
ground state has been depleted:

\begin{equation}
	|c_e(t)|^2 = g(\omega_0)\;|\Omega_{e\leftarrow
			g}|^2\;t,
\label{rate_equation}
\end{equation}

\noindent so that the rate $W_{e\leftarrow g}$ at which the excited-state
population grows due to the applied radiation is:

\begin{equation}
	W_{e\leftarrow g} = g(\omega_0)\;|\Omega_{e\leftarrow g}|^2.
\label{rate}
\end{equation}

\noindent Here $g(\omega_0)$ is the lineshape of the transition (units of
inverse angular frequency, with
$\frac{1}{2\pi}\int_0^\infty\;g(\omega)\,d\omega=1$).\footnote{Even in the
absence of other broadening mechanisms, the finite period of time $\tau$ for
which the perturbative treatment is valid imples a finite Fourier width to the
transition (e.g. $\mathsf{sinc}[(\omega_L-\omega_0)\tau/2]$ in the case of a
square-wave envelope).}

Of course, spontaneous emission \emph{cannot} be ignored.  Even in the absence
of an applied field, the excited state interacts with the vacuum fluctuations
of the electromagnetic field.  The situation here is somewhat different
than that presented above, since the vacuum modes of the electromagnetic
field do \emph{not} represent a narrow-band, directional source.  There are
several approaches to the problem. The most straightforward is a rate-equation
treatment.  Generalizing Eq. (\ref{rate}) to a $\ket{g}\leftarrow\ket{e}$
transition, we must also integrate over all possible modes and sum over the two
orthogonal polarizations possible for each wavevector (directions
$\bm{\epsilon}_1$ and $\bm{\epsilon}_2$).  Now, the number of
plane-wave modes with wavenumbers in the range $[k,\;k+dk]$ in a 
container of volume $V$ is
$dn=\frac{V}{(2\pi)^3}d^3k=\frac{V}{(2\pi c)^3}\omega^2\sin\theta
d\omega d\theta d\varphi$ (in spherical-polar coordinates).  As well, the energy
density corresponding to zero-point energy $\frac{1}{2}\hbar\omega$ in each mode
is $\rho_E(\omega)=\frac{\hbar\omega}{2V}$, and basic electrodynamics tells that
the square of the corresponding electric field
$E_v^2= \frac{\hbar\omega}{2\epsilon_0 V}$.  So, if we denote by $A_{g\leftarrow
e}$ the rate at which the vacuum fluctuations drive population from $|e\rangle$
to $|g\rangle$, then:

\begin{multline}
	A_{g\leftarrow e} = \int_n\sum_{i=1,2}g(\omega) \left|-\frac{\langle 	
g|e\bm{\hat{r}\cdot\epsilon_i}\,E_v|e\rangle}{\hbar}\right|^2 dn \nonumber\\
	= \frac{e^2}{\hbar^2}\int_n\sum_{i=1,2}g(\omega)
|\langle g|\bm{\hat{r}}|e\rangle\bm{\cdot\epsilon}_i|^2 E_v^2 dn \nonumber\\
	= \frac{e^2}{\hbar^2}\int_{\mathbf{k}}\sum_{i=1,2}g(\omega)
|\langle g|\bm{\hat{r}}|e\rangle \bm{\cdot\epsilon}_i|^2
\frac{\hbar\omega}{2\epsilon_0 V} \frac{V}{(2\pi)^3} d^3k \nonumber\\ 
	=\frac{e^2}{2(2\pi c)^3\epsilon_0\hbar}
\int_{\omega}\sum_{i=1,2} g(\omega)|\langle g|\bm{\hat{r}}|e\rangle
\bm{\cdot\epsilon}_i|^2
 \omega^3 \sin\theta d\omega d\theta d\varphi 
\end{multline}

\begin{figure}
\includegraphics[width=63.8 mm,height=46.1 mm]{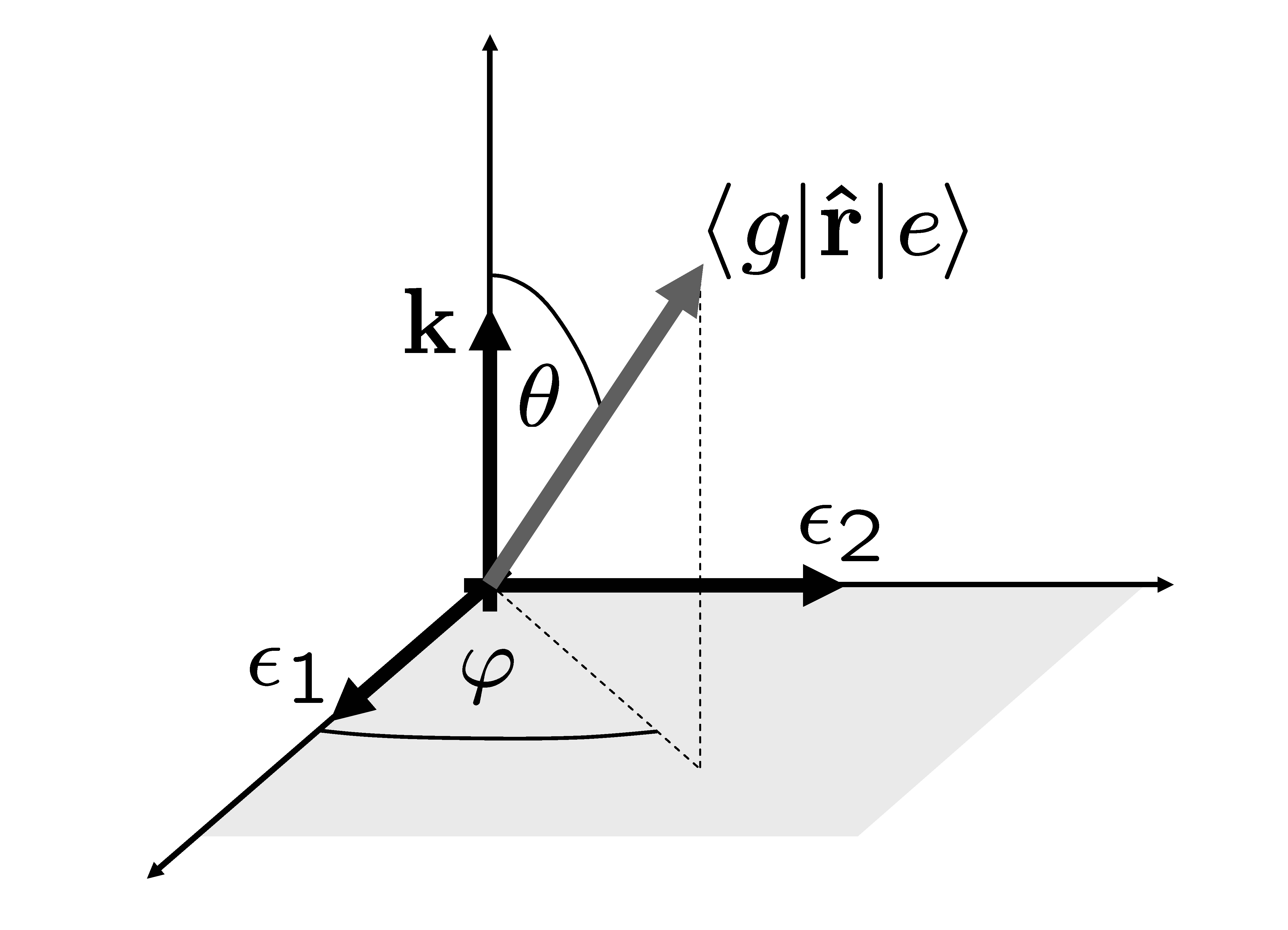}
\caption{Geometry of spontaneous emission:
The $\mathbf{k}$ vector and polarization unit vectors $\bm{\epsilon}_1$ and
$\bm{\epsilon}_2$ associated with the spontaneously emitted photon form an
orthogonal triad with some orientation to the dipole moment
$\bra{g}\mathbf{\hat{r}}\ket{g}$.  We may choose to orient our coordinate axes
aligned with these directions, in which case the dot product
$\bm{\epsilon}_1\cdot\bra{g}\mathbf{\hat{r}}\ket{g}=|\bra{g}\mathbf{\hat{r}}
\ket{ g}|\sin\theta\cos\varphi$ and
$\bm{\epsilon}_2\cdot\bra{g}\mathbf{\hat{r}}\ket{g}=
|\bra{g}\mathbf{\hat{r}}\ket{g}|\sin\theta\sin\varphi$ as the drawing
indicates.}
\label{kr-orientation}
\end{figure}

Now, we must consider geometry. 
$\{\bm{\epsilon}_1,\;\bm{\epsilon}_2,\;\mathbf{k}\}$ form an orthogonal triad,
oriented with respect to $\langle g|\mathbf{\hat{r}}|e \rangle$ as indicated in
Fig. (\ref{kr-orientation}).  Consideration of this diagram indicates that
$\bm{\epsilon}_1\mathbf{\cdot} \langle g|\mathbf{\hat{r}}|e\rangle =
|\langle g|\mathbf{\hat{r}}|e\rangle|\sin\theta \cos\varphi$ and
$\bm{\epsilon}_2\mathbf{\cdot} \langle g|\mathbf{\hat{r}}|e\rangle =
|\langle
g|\mathbf{\hat{r}}|e\rangle|\sin\theta \sin\varphi$ so that

\begin{equation}
	\sum_{i} |\langle g|\mathbf{\hat{r}}|e\rangle\cdot\bm{\epsilon}_i|^2 =
|\langle g|\mathbf{\hat{r}}|e\rangle|^2 \sin^2\theta
\end{equation}

\noindent So finally we have:

\begin{align}
	A_{g\leftarrow e} & = \frac{e^2}{2(2\pi c)^3\epsilon_0\hbar}
\int_{\omega} g(\omega)|\langle g|\bm{\hat{r}}|e\rangle|^2
 \omega^3 \sin^3\theta d\omega d\theta d\varphi \nonumber\\
	& = \frac{e^2}{2(2\pi c)^3\epsilon_0\hbar}\frac{8\pi}{3}
\int_{\omega} g(\omega)|\langle g|\bm{\hat{r}}|e\rangle|^2
 \omega^3 d\omega \nonumber\\
	& = \frac{e^2}{6 \pi^2 c^3\epsilon_0\hbar}
\int_{\omega} g(\omega)|\langle g|\bm{\hat{r}}|e\rangle|^2
 \omega^3 d\omega.
\end{align}

We make the reasonable assumption that the function $g(\omega)$ is sharply
peaked around $\omega_0$ (which is true in practice), so that $\int g(\omega)
|\langle g|\bm{\hat{r}}|e\rangle|^2 \omega^3 d\omega \approx
\omega_0^3\;|\langle
g|\bm{\hat{r}}|e\rangle|^2\;\int g(\omega)d\omega = 2\pi|\langle
g|\bm{\hat{r}}|e\rangle|^2\omega_0^3$.  Finally, we have that

\begin{equation}
\label{A}
	A_{g\leftarrow e} = \frac{e^2\omega_0^3}{3\pi\epsilon_0 \hbar
c^3}|\langle
g|\bm{\hat{r}}|e\rangle|^2 = \frac{8\pi^2e^2}{3\epsilon_0\hbar\lambda_0^3}
|\langle g|\bm{\hat{r}}|e\rangle|^2.
\end{equation}

\noindent
Note that
$\bra{g}\bm{\hat{r}}\ket{e}=\bra{e}\bm{\hat{r}}\ket{g}^*$.

For typical optical dipole-allowed transitions,
$A\sim2\pi\times10^7\;Hz$.\footnote{In units natural to the problem, we can
express the A-coefficient as $A_{g\leftarrow e} = \alpha^3 (R_\infty c)\;
\left(\frac{1}{R_\infty \lambda}\right)^3 |\bra{g}\bm{\hat{\Re}}\ket{e}|^2
\approx (2\pi\times1.278\;GHz) \left(\frac{91.13\;nm}{\lambda}\right)^3
|\bra{g}\bm{\hat{\Re}}\ket {e}|^2$.  Here $\alpha$ is the fine-structure
constant giving the fundamental coupling strength between charged matter and
electromagnetic fields, $R_\infty$ is the Rydberg constant, and
$\bm{\hat{\Re}}=\frac{\mathbf{\hat{r}}}{a_0}$ is the position operator in
units of Bohr radii $a_0$.} We may neglect spontaneous emission (recovering the
Rabi-flopping behaviour described by Eqs. (\ref{Rabi_flopping})) if
$\Omega_{e\leftarrow g}\gg A$.  However, this requires very high laser
intensities.  Although spontaneous emission is driven by only a ``half photon''
in each vacuum mode, there are an \emph{immense} number of such modes in
three-dimensional space.  Thus, it requires a large number of photons in a
\emph{single} (laser) mode to change population at a rate approaching that of
spontaneous emission.  However, if the the ground and excited state are
separated by energies corresponding to long-wavelength, radio-frequency photons,
or if the coupling between them is due to higher-order transitions (electric
quadrupole or magnetic dipole), then it may be possible to realize Rabi
flopping.

A rate-equation treatment of the above type was first performed by
Einstein.\cite{Einstein_1917}  For this reason, the rate $A_{g\leftarrow e}$ is
called the ``Einstein $A$-coefficient.''  In the absence of other broadening
mechanisms (e.g. Doppler or pressure broadening), it gives the natural lifetime
$\tau$ of the excited state and hence the full-width-half maximum
$\Gamma=2\pi\times\Delta\nu$ of the lineshape:

\begin{equation}
	A_{g\leftarrow e} = \Gamma = 2\pi\times\Delta\nu = 1/\tau 
\end{equation}

\noindent (for a two-level atom).  To be explicit, $\Gamma$ is in radians per
second, whereas $\Delta\nu$ is in Hertz.  The lifetime depends only on the
dipole moment of the transition between the levels in question (which goes into
any Rabi frequency calculation) and the energy-density correspoinding to the
zero-point fluctutations of the elctromagnetic field (fixed for our universe).

In principle, given the wavefunctions corresponding to $|g\rangle$ and
$|e\rangle$, we can calculate $\langle g|\mathbf{\hat{r}}|e\rangle$.  However,
\emph{in practice} one only knows the wavefunctions for the hydrogen atom! 
Therefore, we have to rely either on approximate and/or numerical
calculations, or on \emph{measured} quantities such as the lifetime,
and determine $\langle g|\mathbf{\hat{r}}|e\rangle$ using Eq. (\ref{A}).

The NIST database \cite{NISTlevels} of atomic lines lists the appropriate
Einstein-A coefficients for its various lines.  Other databases cite other
quantities such as ``oscillator strengths (f),'' ``cross-sections ($\sigma$),''
or ``line strengths (S).''  I shall not go into the
various definitions and relationships here (see \cite{Cowan_book,
Weissbluth, hilborn_fvalues, Sobelman}).

\subsection{Degeneracy}
\label{degeneracy}

\emph{``There are no two-level atoms...'' - Bill Phillips}

Of course, there \emph{are} no two-level atoms.  However, as long as the two
energy levels in question are distinguishable from other levels (through
frequency or laser polarization, for example), the transition may be treated
\emph{as if} the atom had only the two, aside from ``counting issues'' due to
degeneracy. 

For a simple overview to the changes wrought by degeneracy, let us ignore the
details by which the degeneracy arises, and simply assume that the level we had
called $g$ is, in fact $\mathsf{g}_g$-fold degenerate.  The existence of
multiple ground-state levels implies the possibility of decay into several of
these levels.  (In practice, further physics such as selection rules may
preclude some of the possibilities.)  If we denote the total decay rate by
$A_{g\leftarrow e}$, then:

\begin{equation}
	A_{g\leftarrow e} = \sum_{i=1}^{\mathsf{g}_g}\;A_{g_i\leftarrow e},
\label{1deg:decay}
\end{equation}

\noindent where $A_{g_i\leftarrow e}$ is the decay rate from $e$ to the $i$-th
sublevel.

Suppose now that the excited state $e$ is also degenerate, having a
$\mathsf{g}_e$-fold degeneracy.  Several points may be made.  First, from a
thermodynamic point of view, we must demand (and the physics will deliver!)
that the \emph{total} decay $A_{e_j}=\sum_{i=1}^{\mathsf{g}_g}A_{g_i\leftarrow
e_j}$ from each upper sublevel $e_j$ be equal;  if this were not the case, then
thermal excition of the excited state would result in unequal
steady-state population of the excited state (due to the unequal decay rates). 
We will see below how this equal total decay rate arises in the case where the
degeneracy is due to angular momentum.  Second, in a case where the degenerate
excited state sublevels are populated with probabilities $\mathcal{P}_{e_j}$,
then the total decay rate measured is the \emph{average} of the decay rates of
each sublevel (each of which can possibly decay to multiple ground-state
sublevels).

\begin{equation}
	A_{g\leftarrow e,
distrib.}=\sum_{j=1}^{\mathsf{g}_e}{\mathcal{P}_{e_j}}\;
\sum_{i=1}^{\mathsf{g}_g}A_{g_i\leftarrow e_j}.
\end{equation}

\noindent For a thermally populated excited state the
probabilities $\mathcal{P}_{e_j} = 1/g_e$ are equal.  This is the case for,
e.g., the distribution produced by the discharge lamps historically used for
atomic spectroscopy. This distribution may or may not be relevant to more modern
spectroscopic measurements. (However, perhaps for historical reasons, it is
ubiquitous in books on atomic spectroscopy.)

\section{Overview of Rotational Symmetry and Angular Momentum}
\label{symmetry}

In standard atomic systems, degeneracies inevitably arise from angular
momentum considerations.  Before considering the physics and math behind this
degeneracy, it will pay to briefly review rotational symmetry and angular
momentum in quantum systems.  Symmetry plays a powerful role in classical
mechanics, as epitomized by Noether's theorem.\cite{Hamel:1912, Noether,
Marion_Thornton, Landau_Lifshitz_Mech}  However, in classical mechanics,
invariance of the equations of motion does not necessarily imply symmetry of a
motional state.  In quantum systems, on the other hand, superposition implies
that the quantum states themselves may always be expressed so as to reflect the
symmetries of the underlying Hamiltonian.\cite{Wigner_symmetry,
Wigner_grouptheory, Weyl, Racah_1, Racah_2, Racah_3, Racah_4, Thompson_L} 

This has far-reaching implications for atomic physics, where the spherically
symmetric Coulomb potential dominates the physics.  So let us consider
rotational symmetry.  From a purely geometric point of view, the operators

\begin{align}
\label{geo_generators}
 \mathcal{L}_x &= i\left(\sin\varphi\frac{\partial}{\partial\theta} +
		\cot\theta \cos\varphi\frac{\partial}{\partial\varphi}\right)\\
 \mathcal{L}_y &= i\left(-\cos\varphi\frac{\partial}{\partial\theta} +
		\cot\theta \sin\varphi\frac{\partial}{\partial\varphi}\right)\\
 \mathcal{L}_z &= -i\frac{\partial}{\partial\varphi}
\end{align}

\noindent
generate rotations of a function $f(x,y,z)$ of the spatial coordinates $x$,
$y$, and $z$.  That is, if we rotate the function $f$ an angle $\theta$
about the axis $\mathbf{n}$, then
$f'(x,y,z)=\mathcal{R}f(x,y,z)=e^{-i\theta\bm{n\cdot\mathcal{L}}}f(x,y,z)$,
where $\mathcal{R}$ represents a rotation operator.\footnote{If
$\mathcal{R}_x(\theta)$, $\mathcal{R}_y(\theta)$, and $\mathcal{R}_z(\theta)$
represent the rotation operator for a rotation about the original $x$, $y$, and
$z$ axes by angle $\theta$ respectively , then we take the Euler angles
$\alpha$, $\beta$, $\gamma$ to be such that
$e^{-i\theta\bm{n\cdot\mathcal{L}}}f(x,y,z)=\mathcal{R}_z(\alpha)\mathcal{R}_y
(\beta)\mathcal{ R}_z(\gamma) f(x,y,z)$.  This is the convention followed
by Messiah.\cite{Messiah}}  This is the so-called \emph{active} view of
rotations, where we change the function while holding our coordinate axes fixed.

In quantum mechanics, deBroglie's fundamental relation $\hat{p}=-i\hbar\nabla$
gives the quantities $\hat{L}_k=i\hbar{\mathcal{L}}_k$ not just
\emph{geometrical} significance but also \emph{dynamical} and, by the
postulates of quantum mechanics, \emph{observable} consequences as components
of \textbf{angular momentum} (e.g. the quantized outcome of the measurements of
angular momentum projections $\hat{L}_z$).\footnote{The distinction and
commonality of the geometrical generators of rotations and the dynamical,
quantum angular momentum components is discussed by Dirac,\cite{Dirac_text}
Wigner,\cite{Wigner_symmetry, Wigner_grouptheory} and
Thompson,\cite{Thompson_L} amongst others.}

Because rotations about different axes do not commute, the operators
$\mathcal{L}_i$ obey the commutation relations
$\left[\mathcal{L}_i,\mathcal{L}_j\right]=i\epsilon_{ijk}\mathcal{L}_k$ or, more
compactly,
$\bm{\mathcal{L}\times\mathcal{L}}=i\bm{\mathcal{L}}$.  In the quantum case, the
quantum mechanical angular momentum operators $\hat{L}_k$ obey the related
commutation relations
$\left[\hat{L}_i,\hat{L}_j\right]=i\hbar\epsilon_{ijk}\hat{L}_k$, or
$\mathbf{\hat{L}\times\hat{L}}=i\hbar\mathbf{\hat{L}}$.  These commutation
relations identify a general quantum mechanical operator $\mathbf{\hat{J}}$
as being an angular momentum.

Given the fact of that rotations about different axes do not commute, let us
focus on only a single axis of rotation - which we will call the $z$ axis - and
ask which directions in space are invariant under rotations about this axis. 
The eigenvectors of the rotation operator
$\mathcal{R}(\varphi,\;\mathbf{e}_{z})$ are given by: \cite{Biedenharn_Louck1}
\begin{eqnarray}
	\mathcal{R}(\varphi,\;\mathbf{e}_z)\;(\mathbf{e}_x + i\,\mathbf{e}_y)
&=&
e^{-i\varphi}\;(\mathbf{e}_x + i\,\mathbf{e}_y)\nonumber\\
	\mathcal{R}(\varphi,\;\mathbf{e}_z)\;(\mathbf{e}_x - i\,\mathbf{e}_y)
&=&
e^{i\varphi}\;(\mathbf{e}_x - i\,\mathbf{e}_y)\nonumber\\
	\mathcal{R}(\varphi,\;\mathbf{e}_z)\;\mathbf{e}_z &=& \mathbf{e}_z
\end{eqnarray}
\noindent Of course, the first two eigen``vectors'' are not physical vectors at
all, since they're complex.  Normally, realizing that we are talking about
real, three-dimensional space, we would ``toss out'' these solutions.  However,
it turns out that these vectors have physical use after all.  Indeed, 
they may seem somewhat similar to the definition of quantum angular
momentum raising/lowering operators $\hat{L}_\pm:=\hat{L}_x\pm i\;\hat{L}_y$
or the ``spherical basis unit vectors'' $\mathbf{e}_{\pm 1}:=
\mp\frac{1}{\sqrt{2}}\left(\mathbf{e}_x \pm
i\,\mathbf{e}_y\right)$, which are proportional to the eigenvectors. 
This is no coincidence - these entities are useful exactly because of their
similarity to the expressions for the eigenvectors of rotation.  Due to
the vectors' simple rotational properties, they are particularly useful in
describing changes in a physical system induced by rotations.  Since
quantum-mechanical wave functions are delocalized and complex anyway, the
complex-valued unit vectors prove useful in describing quantum systems.

However, the complex nature of $\unit{\pm 1}$ requires some notational caution,
since we must ensure that quantities with a real, physical meaning - such as the
dot product $\mathbf{A}\cdot\mathbf{B}$ of two real vectors - evaluates to a
real number.  One way to assure this is to expand our vector notation by
introducing \emph{dual vectors}:  this is the approach which gives us
\emph{bras} $\bra{\Psi}$ (dual vectors), and \emph{kets} $\ket{\Psi}$ (state
vectors) in quantum mechanics. A similar rationale gives us \emph{contravariant}
vector components $\mathcal{A}^\mu$ and \emph{covariant} dual vector components
$\mathcal{A}_\mu$ in relativity.  Given a vector $\mathbf{A}=A_x\,\unit{x} +
A_y\,\unit{y} + A_z\,\unit{z}$, we define:

\begin{subequations}
\begin{align}
 	\Unit{+1} & := -\frac{1}{\sqrt{2}}\left(\unit{x} -
i\,\unit{y}\right)\\
	\Unit{0} & := \unit{z}\\
	\Unit{-1} & := +\frac{1}{\sqrt{2}}\left(\unit{x} + i\,\unit{y}\right),
\end{align}
\end{subequations}

and

\begin{subequations}
\begin{align}
 	A_{+1} & := -\frac{1}{\sqrt{2}}\left(A_x + i\,A_y\right)\\
	A_0 & := A_z\\
	A_{-1} & := +\frac{1}{\sqrt{2}}\left(A_x - i\,A_y\right),
\end{align}
\end{subequations}

\noindent
and let $A^q=(A_q)^*$ and $\unit{q}=(\Unit{q})^*$ (where
$q\in\left\{-1,0,+1\right\}$).  Really, the notation is just a way of keeping
track of complex conjugation, but it is consistent with other notations the
reader may be familiar with, and also is consistent with various notations in
the literature.  In terms of these quantities, we may express $\mathbf{A}$ as:

\begin{eqnarray}
 	\mathbf{A} & = & A_x\,\unit{x} + A_y\,\unit{y} + A_z\,\unit{z}
\nonumber\\
		& = & A_{+1}\,\Unit{+1} + A_0\,\Unit{0} + A_{-1}\,\Unit{-1}
\end{eqnarray}

\noindent
More compactly, $\mathbf{A} = \sum_q A_q\,\Unit{q}$.

As an example, we may express a general position vector $\mathbf{r}$ as

\begin{align}
\label{spherical_position}
 \mathbf{r}&=\sqrt{\frac{4\pi}{3}}r\left[Y_{-1}^1\,\Unit{-1} + Y_0^0\,\Unit{0}
		+Y_{+1}^1\,\Unit{+1}\right]\nonumber\\
	&=r\left[C_{-1}^1\,\Unit{-1}+C_0^0\,\Unit{0}+C_{+1}^1\,\Unit{+1}\right],
\end{align}

\noindent
where $C_m^l:=\sqrt{\frac{4\pi}{2l+1}}Y_m^l$ are the ``normalized spherical
harmonics'' introduced by Racah, which save us writing inumerable factors
of $\sqrt{\frac{4\pi}{2l+1}}$.  Note that, if the expansion coefficients are in
fact to be equal to the usual spherical harmonics,\cite{Messiah} then we
\emph{must} write the expansion in the above form, using the unit vectors
$\Unit{m}$.  This implies that the spherical harmonics transform as
``covariant'' quantities in this notation.

In terms of these definitions, the dot product of two vectors $\mathbf{A}$ and
$\mathbf{B}$ is given by $\mathbf{A}\cdot\mathbf{B} = \sum_q A^qB_q = \sum_q
A_q^*B_q = \sum_q (-1)^q\,A_{-q}B_q$.  Note that
$\Unit{q}\cdot\unit{r}\equiv\unit{q}^*\cdot\unit{r} = \delta_{q,r}$.  The
multiplicity of equivalent expressions may seem daunting, but the reader will
find all of them in the literature, so I have included them here.  I will stick
to notation such as $\mathbf{A}\cdot\mathbf{B}=\sum_q A^qB_q$.

\subsection{Coupled angular momenta:  Clebsch-Gordon and n-j Symbols}
\label{coupling}

If we combine two states with definite rotational symmetry (i.e. angular
momentum eigenstates), then the resulting state will reflect these symmetries. 
Consider, for example, a single outer electron in an atom.  The electron has
both orbital angular momentum
$\mathbf{\hat{L}}$ and spin $\mathbf{\hat{S}}$, with quantum numbers $l$,
$m_l$ and $s$, $m_s$, respectively.  The components of these angular momenta
satisfy the usual commutation relations. However, the combined system has
angular momentum $\mathbf{\hat{J}}=\mathbf{\hat{L}}+\mathbf{\hat{S}}$ with
quantum numbers $j$ and $m$. The combined system can be expressed either in
terms of the state vectors $\ket{lm_lsm_s}$ or in terms of the state
vectors $\ket{lsjm}$.  The two choices are consistent - we can write the states
$\ket{lsjm}$ in terms of the states $\ket{lm_lsm_s}$:

\begin{equation}
\label{J_expansion}
 	\ket{lsjm}=\sum_{m_l,m_s}C_{m_lm_sm}^{lsj}\ket{lm_lsm_s}.
\end{equation}

\noindent
Here, the expansion coefficients $C_{m_lm_sm}^{lsj}$ are the
\emph{Clebsch-Gordon coefficients}:

\begin{equation}
\label{CG}
 	C_{m_lm_sm}^{lsj} = \langle lm_lsm_s|lsjm\rangle.
\end{equation}

The reader has no doubt encountered the Clebsch-Gordon coefficients before. 
They simply represent overlap between the state $\ket{lm_lsm_s}$
and the state $\ket{lsjm}$ - that is to say, the ``amount'' of $\ket{lm_lsm_s}$
``in'' the state $\ket{lsjm}$.

However, in performing angular-momentum calculations, it is usually more
convenient to introduce the \emph{Wigner 3-j symbols}:

\begin{equation}
\label{3j}
	\Threej{l}{s}{j}{m_l}{m_s}{m} =
\frac{(-1)^{l-s-m}}{\sqrt{2j+1}}\langle lm_lsm_s|lsj-m\rangle.
\end{equation}

\noindent
\textbf{By convention, the Clebsch-Gordon coefficients are taken to be real,.
  Thus, the 3-j symbols are also real (positive or negative) numbers.}  The 3-j
symbols exhibit a number of simple relationships with other 3-j symbols where
the arguments are permuted.  An even permutation of symbols leaves the 3-j
symbol unchanged:

\begin{equation}
\label{3j_evenperm}
 	\Threej{l}{s}{j}{m_l}{m_s}{m} = \Threej{j}{l}{s}{m}{m_l}{m_s} =
\Threej{s}{j}{l}{m_s}{m}{m_l}
\end{equation}

\noindent
whereas an odd permutation introduces only a phase factor:

\begin{equation}
\label{3j_oddperm}
 	(-1)^{l+s+j}\Threej{l}{s}{j}{m_l}{m_s}{m}
	 = \Threej{s}{l}{j}{m_s}{m_l}{m},\;etc.
\end{equation}

\noindent
Finally, we have the relationship
\begin{equation}
 \label{3j_flip}
 	\Threej{l}{s}{j}{m_l}{m_s}{m} =
(-1)^{l+s+j}\Threej{l}{s}{j}{-m_l}{-m_s}{-m}
\end{equation}

\noindent
There are similar relationships between Clebsch-Gordon coefficients, but
these relationships are encumbered by various factors of $\sqrt{2j+1}$, etc.,
and are less wieldy to work with.

Roughly speaking, the 3-j symbol gives the probability (amplitude) that angular
momentum $l$ with projection $m_l$ will add up with an angular momentum $s$ with
projection $m_s$ to produce an angular momentum $j$ with projection $-m$ - but
normalized to the \emph{total number} $2j+1$ of possible distinct orientations
of $j$. This choice of normalization is responsible for the $\sqrt{2j+1}$ in the
denominator on the right side of Eq. (\ref{3j}), and is necessary for the
convenient permutational symmetries of the 3-j symbols to hold.  There is
another way to interpret the 3-j symbols, as corresponding to the
probability (amplitude) that if one adds an angular momentum $\mathbf{\hat{L}}$
(with projection $m_l$) and an angular momentum $\mathbf{\hat{S}}$ (with
projection $m_s$) and then \emph{subtracts} an angular momentum
$\mathbf{\hat{J}}$ (with projection $m$ so that $-\mathbf{\hat{J}}$ has
projection $-m$), one obtains an angular momentum $0$ - that is,
a scalar (rotationally invariant) quantity.  Physically, this simply reflects
the fact that in a system that conserves angular momentum, angular momentum is
conserved!  The probability is again normalized to the total number $2j+1$ of
angular momentum states $j$.

The 3-j symbols arise in combining two angular momenta to make a third (or,
alternatively, coupling 3 angular momenta to form a $j=0$ scalar state). 
Similar considerations arise in combining 3 angular momenta.  Consider angular
momenta $j_1,\;m_1$, $j_2,\;m_2$, and $j_3,\;m_3$ which we combine to form an
overall angular momentum $j,\;m$.  We can do this by first coupling $j_1$
and $j_2$ to form an angular momentum eigenstate $j_{12}$ (with projection
$m_{12}$), and then couple $j_{12}$ with $j_3$ to obtain the state
$\ket{(j_1j_2)j_{12}j_3jm}$. However, we can \emph{also} first couple $j_2$ and
$j_3$ to form an angular momentum $j_{23}$ (with projection $m_{23}$), and then
combine $j_1$ with $j_{23}$ to form $\ket{j_1(j_2j_3)j_{23}jm}$.  Either scheme
is appropriate - however, the two kets $\ket{(j_1j_2)j_{12}j_3jm}$ and
$\ket{j_1(j_2j_3)j_{23}jm}$ are not, in general, the same.  Nonetheless, we can 
expand the state $\ket{(j_1j_2)j_{12}j_3jm}$ in terms of the various states
$\ket{j_1(j_2j_3)j_{23}jm}$:

\begin{widetext}
\begin{equation}
 	\ket{(j_1j_2)j_{12}j_3jm}=\sum_{j_{23}}\langle
j_1(j_2j_3)j_{23}jm|(j_1j_2)j_{12}j_3jm\rangle\;|j_1(j_2j_3)j_{23}jm\rangle.
\end{equation}
\noindent
The \emph{Wigner 6-j} symbol is defined as:

\begin{equation}
\noindent
	\Sixj{j_1}{j_2}{j_{12}}{j_3}{j}{j_{23}}=\frac{(-1)^{j_1+j_2+j_3+j}}
{\sqrt{(2j_{12}+1)(2j_{23}+1)}}\langle
j_1(j_2j_3)j_{23}jm|(j_1j_2)j_{12}j_3jm\rangle.
\end{equation}
\end{widetext}

\noindent
6-j symbols are a notationally convenient way of keeping track of the coupling
between 3 angular momenta.  Similarly to case of the 3-j symbols, one may
interpret the 6-j symbols in terms of adding 3 angular momenta, and subtracting
a fourth to obtain a $j=0$ scalar.  The 6-j symbol has the nice symmetry that
its value is unchanged by the interchange of any two of the three columns, or by
switching the upper and lower members of any two columns.

To obtain some insight as to the meaning of a 6-j symbol, consider the
quantity $\sixj{s}{l}{j}{1}{j'}{l'}$.  In terms of the
definition

\begin{align}
	\sixj{s}{l}{j}{1}{j'}{l'}=\frac{(-1)^{s+l+1+j'}}{\sqrt{(2j+1)(2l'+1)}}
	\langle s(l1)l'j'm\ket{(sl)j1j'm},
\end{align}

\noindent
we see that the 6-j symbol is proportional to the overlap between two
states.  The first is one in which the initial orbital angular momentum $l$ is
first coupled to the unit angular momentum of the laser field to form
the new angular momentum $l'$, which is then in turn coupled to the original
spin $s$ (which is unaffected by the laser!) to form the final total angular
momentum $j'$.  The second state is one in which spin $s$ is first coupled to
orbital angular momentum $l$ to form total atomic angular momentum $j$, and then
$j$ is coupled to the unit angular momentum of the laser field to form total
angular momentum $j'$ - the angular momentum of the final state.  So essentially
the 6-j symbol is
related to the two different ways of thinking about the atom-laser coupling: 
either as affecting the \emph{total} angular momentum of the atom, or as
affecting only its \emph{orbital} angular momentum.  The factor of
$1/\sqrt{(2j+1)(2l'+1)}$ normalizes to the product of the total numbers of
intermediate states, and is necessary for the 6-j symbols' permutation
symmetries.

One may also introduce 9-j symbols, etc., but I promise the reader that I
will not do so here!

\subsection{Introduction to the \emph{Wigner-Eckart Theorem}}
\label{Wigner_Eckart}

The entire reason for introducing the whole apparatus of the previous pages is
that the notation makes explicit the symmetry of states, vectors, operators,
etc. under rotations.  Thus, the language is well-suited to describing systems
that exhibit rotational symmetry.  This symmetry can save us an immense amount
of work if we make use of it, and the notation allows this.

The greatest implication of rotational symmetry is embodied in the
\emph{Wigner-Eckart theorem}.\cite{Messiah}  Suppose that we have two states of
well-defined rotational symmetry, and some physical interaction that also
exhibits a well-defined rotational symmetry couples the two states.  To
rephrase, suppose that two angular-momentum eigenstates $\ket{\alpha j m}$ and
$\ket{\alpha' j'm'}$ are coupled by an irreducible tensor operator
$\mathbf{\mathsf{T}^{(k)}}$ with components $T_q^k$ (see Refs.
\cite{Messiah,Butkov,Arfken_Weber,Silver_tensors,Suhonen}). Here, the
labels $\alpha$, $\alpha'$ represent any additional labels in addition to
angular momentum needed to uniquely specify the quantum states.  For example, in
describing the orbital of a hydrogen atom, one would need to specify the
principal quantum number $n$.  The matrix element for the transition is $\langle
\alpha' j'm'|T_q^k|\alpha j m\rangle$.  However, since each term in the matrix
element has well-defined rotational symmetry, so too must the overall matrix
element.  To put it in more active terms (in view of the quantum relationship
between generators of rotations and angular momentum), angular momentum is
conserved in the transition.

The Wigner-Eckart theorem essentially splits the calculation of the matrix
element into a term that embodies the peculiar specifics of the particular
interaction and a term that embodies the \emph{purely geometric} considerations
demanded by the rotational symmetry - that is, by conservation of angular
momentum.  To be quantitative, the Wigner-Eckart theorem states
that\footnote{\label{WE_note}As Silver points out,\cite{Silver_tensors}
conventions for expressing the Wigner-Eckart theorem group various minus signs
and factors of $\sqrt{2j+1}$ with the reduced matrix element.}

\begin{widetext}
\begin{equation}
\label{WE_eq}
 	\langle \alpha' j' m'|T_q^k|\alpha j m\rangle = (-1)^{j'-m'} 
\langle\alpha' j'||\mathbf{\mathsf{T}}^{(k)}||\alpha j\rangle
\Threej{j'}{k}{j}{-m'}{q}{m}.
\end{equation}
\end{widetext}

Note that the \emph{reduced matrix element} (or \emph{double-bar matrix
element}) is a constant independent of the quantum numbers $m_j$, $m_j'$, and
$q$.  That is to say, $\langle \alpha' j'||\mathbf{\mathsf{T}}^{(k)}||\alpha
j\rangle$ is the same regardless of the relative orientations of the angular
momenta $j$, $j'$, and $k$ (the angular momentum associated with the operator).
$\langle\alpha' j'||\mathbf{\mathsf{T}}^{(k)}||\alpha j\rangle$ expresses the
physics of the particular interaction at hand - and, as such, it \emph{does}
contain information about the angular momenta of the initial and final states
and the effective angular momentum of the interaction driving transitions
between these states.  However, the dependence of the transition strength on
the relative orientation of the rotationally symmetric quantities is a question
of \emph{pure geometry} given the well-characterized rotational symmetries of
the quantities involved.  \emph{It is entirely independent of the details of
the interaction} and the same for \emph{any} transition between angular
momentum eigenstates driven by an interaction with the rotational symmetry
characteristic of angular momentum $k$.  This universal geometric part of the
transition matrix element is given by the factor of
$(-1)^{j'-m'}\threej{j'}{k}{j}{-m'}{q}{m}$.

The practical upshot of the Wigner-Eckart theorem is that the transition matrix
elements for a particular coupling between angular momentum eignenstates $j$,
$j'$ \emph{is the same for all the states} - up to a multiplicative geometric
factor which factors in relative orientations.  This geometric factor
$(-1)^{j'-m'}\threej{j'}{k}{j}{-m'}{q}{m}$ (which may be zero!) can be
looked up in standard tables or computed with standard software packages.  The
reduced matrix element, on the other hand, describes the actual specific
physics at hand, and must be calculated explicitly for each physical setup.

\section{Rabi frequencies for an atom with spin and orbital angular momentum}
\label{J}

After the long digression on angular momentum, let us return to the question of
the Rabi frequency.  Our digression has equipped us with the tools to calculate
the transition strength with a minimum of tedium.

For an atom with a single outer electron (ground s-state), consider laser-driven
electric-dipole transitions between states $\ket{njm}$ and $\ket{n'j'm'}$. 
Here, $j$ ($j'$) is the vector sum of the electron's orbital angular momentum
(the angular variation of the electron's wave function) and the electron spin. 
However, in the electric dipole approximation, the electric field of the the
laser does not couple to the  electron spin. So, if you will, the electric field
couples only to the ``$l$ ($l'$) part'' of $j$ ($j'$).  (Note that in the rest
of the paper, I will neglect fine-structure, hyperfine-structure and Zeeman
splittings, in order to focus on the essential commonality of the various
transitions.)

One way to calculate the Rabi frequency, then, would be to decompose $j$
into $l$ and $s$, and evaluate the transition matrix element between different
eigenstates of $\hat{L}^2$, $\hat{L}_z$, with the electronic spin being
``carried along for the ride.'' This is the approach suggested in Ref.
\cite{CoolingNeutrals}.

However, the Wigner-Eckart theorem offers us a simpler approach - particularly
if we wish to calculate the Rabi frequency in terms of the excited-state
lifetime.  The point is that it doesn't matter
\emph{how} the angular momentum $j$ arises.  It only matters that the initial
and final states are states of well-defined rotational symmetry (angular
momentum) and that the interaction potential may be expressed in a similar
manner.

In particular, we have that $\hat{V}_I = -\bm{\hat{\mu}}\cdot\mathbf{E}$. 
In order to evaluate the dot product, we have to pick a coordinate system.  We
know from the quantum theory of angular momentum that only one component of
$\mathbf{\hat{J}}$ can have a well-defined value, and by convention, we call
that direction the $z$ direction.  Now, an isolated atom has spherical symmetry,
and by that token, \emph{it does not matter which direction we choose to call
the} $z$-\emph{direction}.  However, in practice, the perfect spherical symmetry
is broken by some outside perturbation.  In typical atom-trapping experiments,
this is provided by a uniform applied magnetic field - referred to as the
``quantization field.''  The magnetic field ``picks out'' a ``preferred
direction'' in space and breaks the degeneracy of the different atomic states
through the well-known Zeeman effect.  In this case, it is wise to pick as the
$z$-axis the axis of this background field.\footnote{In the absence of a
background magnetic field, an \emph{unambiguous} choice for the $z$-axis is the
direction $\mathbf{k}$ of the beam's propagation.}  We need not worry about the
particular directions of $x$ and $y$ for we shall calculate in the spherical
basis $\unit{+1}$, $\unit{0}$, $\unit{-1}$.

Once we have picked a $z$, or quantization, axis we can then express the
laser electric field components in that basis.  By convention, a laser field
(component) parallel to the $z$ axis is said to have ``$\pi$ polarization.'' 
A laser field which, in the rotating-wave approximation, drives a lower level
$\ket{njm}$ to an upper level $\ket{n'j'(m+1)}$ is said to have ``$\sigma+$
polarization'' and a laser field which drives a lower level $\ket{njm}$ to an
upper level $\ket{n'j'(m-1)}$ is said to have ``$\sigma-$ polarization.''  In
considering such a transition, a $\sigma+$ ($\sigma-$) field would, in the
rotating-wave approximation, have an electric field with only a $\unit{+1}$
($\unit{-1}$) component.\footnote{In this scheme the \emph{only} way for the
laser to be $\pi$-polarized is if the electric field is linearly polarized
\emph{and parallel to the $z$ axis} That is, the $\mathbf{k}$ vector of the
laser must be perpendicular to the quantization axis.  If the laser's electric
field is \emph{not} parallel to the quantization axis, then the laser will have
$\sigma+$ and $\sigma-$ components \emph{even if the field is linearly
polarized}.}.

The atom's dipole moment is given by $\mathbf{\hat{\mu}}=-e\mathbf{\hat{r}}$. 
Using Eq. (\ref{spherical_position}),
$\mathbf{\hat{r}}=\hat{r}\sum_{q}C_q^1\,\Unit{q}$.  In terms of the above
expressions:

\begin{equation}
 	\hat{V}_I=-\pmb{\hat{\mu}\cdot{E}} = e\sum_q \epsilon^q \hat{r}\,C_q^1.
\end{equation}

\noindent
This finally expresses the interaction Hamiltonian in a way which brings to the
forefront the rotational symmetry of the situation and which, more
significantly, allows us to calculate the Rabi frequency using the
Wigner-Eckart theorem.

The Rabi frequency is given by:

\begin{align}
 	\Omega_{g\leftarrow e} &= \frac{1}{\hbar}\langle n'j'm'|eE_0\sum_q
\epsilon^q \hat{r} C_q^1|njm\rangle\nonumber\\
		& = \frac{eE_0}{\hbar}\sum_q \epsilon^q \langle
n'j'm'|\hat{r}C_q^1 |njm\rangle.
\end{align}

\noindent
Now $\hat{r}$ is an isotropic (scalar) operator, which has no effect in the
space $\ket{jm}$.  Thus, the transformation properties of the constituents in
the sum above will be set by the angular momentum eigenstates and the operators
$C_q^1$ $\propto{Y_q^1}$.  But here the Wigner-Eckart theorem simplifies life,
for it assures us that, regardless of the values of $j,m,j',m',q$:

\begin{equation}
 	\langle n'j'm'|\hat{r}C_q^1|njm\rangle =(-1)^{j'-m'}\langle n'j'||
\hat{r}\mathbf{\mathsf{C}^{(1)}}||nj\rangle\Threej{j'}{1}{j}{-m'}{q}{m}.
\end{equation}

The $3-j$ symbols may be looked up in tables or calculated, and the
so-called \emph{reduced matrix element} $\langle n'j'm'||
\hat{r}\mathbf{\mathsf{C}^{(1)}}||njm\rangle$ is independent of the
various projection quantum numbers.  (The symbol $\mathbf{\mathsf{C}^{(1)}}$
represents the first-order tensor of which the $C_q^1$ are components.)  So
finally:

\begin{equation}
\label{3jRabi}
 	\Omega_{e\leftarrow g} 
		 =\frac{eE_0}{\hbar}(-1)^{j'-m'}\langle n'j'||
\hat{r}\mathbf{\mathsf{C}^{(1)}}||nj\rangle \sum_q \epsilon^q
\Threej{j'}{1}{j}{-m'}{q}{m}.
\end{equation}

\noindent
Various selection rules follow from Eq. (\ref{3jRabi}), since the
$3-j$ symbol vanishes unless $j'-j=0,\;\pm 1$, $j'+j\geq 1$, and $m'-m=0,\;\pm
1$.

One interpretation of Eq. \ref{3jRabi} is as follows.  The factor of
$\frac{eE_0}{\hbar}$ and the reduced matrix element express the size of the
dipole moment induced in the atom by the applied electric field of the
laser.\footnote{In fact, the product
$\frac{eE_0}{\hbar}\bra{n'j'}|\hat{r}\mathbf{\mathsf{C}^{(1)}}|\ket{nj}$ is
$\sqrt{2j+1}$ times the dipole moment, due to the normalization of the 3-j
symbols.  That is, the multiplicity of possible excited-state levels
``dilutes'' the transition strength to a given level.} The sum over 3-j symbols
then expresses the relative orientation between the electric field and the
dipole moment of the atom when it is in a superposition of states $\ket{n'j'}$
and $\ket{nj}$.

For laser-cooling experimentalists, our work is now all but over.  For
we can use the lifetime of the excited state to determine the reduced matrix
element in Eq. (\ref{3jRabi}), in the case of an excited state 
$\ket{n'j'm'}$ which can only decay to the manifold $\ket{njm}$ (typical of
$S\rightarrow P$ transitions).  First, recall that Eqs. (\ref{A}) and
(\ref{1deg:decay}) tell us how to calculate the total decay rate from 
the particular excited state $\ket{n'j'm'}$.  Next, note
that$\bra{njm}\hat{r}C_q^1\ket{n'j'm'}=
\bra{n'j'm'}\hat{r}C_q^{1*}\ket{njm}=\bra{n'j'm'}\hat{r}(-1)^q C_{-q}^1\ket{njm}
$. (Basically, this statement reflects the fact that if an \emph{absorbed}
photon increases (decreases) the angular momentum of the atomic state, then an
\emph{emitted} photon must do the converse).  Putting this all together, we
have:

\begin{widetext}
\begin{align}
 	\Gamma &= A_{|nj\rangle \leftarrow |n'j'm'\rangle} =
\sum_{q,m}A_{|njm\rangle \leftarrow |n'j'm'\rangle}\nonumber\\
		& = \frac{8\pi^2e^2}{3\epsilon_0\hbar
\lambda_0^3}\sum_{q,m}|\langle njm|{\hat{r}} C_q^1|n'j'm'\rangle|^2\nonumber \\
& = \frac{8\pi^2e^2}{3\epsilon_0\hbar
\lambda_0^3}\sum_{q,m}|\langle n'j'm'|{\hat{r}} C_{-q}^1|njm\rangle|^2
\nonumber\\
& = \frac{8\pi^2e^2}{3\epsilon_0\hbar
\lambda_0^3}|\langle n'j'||
\hat{r}\mathbf{\mathsf{C}^{(1)}}||nj\rangle|^2\sum_{q,m}
\Threej{j}{1}{j'}{-m}{-q}{m'}\Threej{j}{1}{j'}{-m}{-q}{m'}
\end{align}

Now we  can simplify the sum over squares of 3-j symbols by their 
tabulated properties.  In particular Eq. (C.15a) of Messiah\cite{Messiah} tells
us that

\begin{equation} 
\sum_{m_1=-j_1}^{+j_1}\sum_{m_2=-j_2}^{+j_2}\Threej{j_1}{j_2}{j_3}{m_1}{m_2}{m_3
}\Threej{j_1}{j_2}{j_3'}{m_1}{m_2}{m_3'}=\frac{1}{2j_3+1}\delta_{j_3,j_3'}
\delta_{m_3,m_3'},
\end{equation}
\end{widetext}

\noindent
which, as applied to this case, yields:

\begin{equation}
 	\sum_{q,m}
\Threej{j}{1}{j'}{-m}{-q}{m'}\Threej{j}{1}{j'}{-m}{-q}{m'}=\frac{1}{2j'+1}.
\end{equation}

\noindent
(The fact that the sum evaluates to $1/(2j'+1)$ is a result of the
normalization of the 3-j symbols.)

Thus,

\begin{equation}
\label{J_lifetime}
 	\Gamma = \frac{1}{2j'+1}\frac{8\pi^2e^2}{3\epsilon_0\hbar
\lambda_0^3}|\langle n'j'||\hat{r}\mathbf{\mathsf{C}^{(1)}}||nj\rangle|^2.
\end{equation}

By a systematic and careful comparison with the results of the next section 
(see Appendix \ref{proof}), the phase of the reduced matrix element can be fixed
as $(-1)^{j+j_>}$ (where $j_>$ is the larger of $j'$ and $j$), so that

\begin{equation}
\label{RME_phase}
 	\langle n'j'||\hat{r}\mathbf{\mathsf{C}^{(1)}}||nj\rangle
= (-1)^{j+j_>}\sqrt{2j'+1}\sqrt{\frac{3\epsilon_0\hbar \lambda_0^3
\Gamma}{8\pi^2e^2}}.
\end{equation}

So finally, for a transition whose lifetime is known to be $1/\Gamma$, the Rabi
frequency may be calculated as:

\begin{widetext}
\begin{equation}
\label{actual_Rabi}
	\Omega_{e\leftarrow g} =\frac{E_0}{\hbar}\sqrt{\frac{3\epsilon_0\hbar
\lambda_0^3 \Gamma}{8\pi^2}}(-1)^{j+j'+j_>-m'}\sqrt{2j'+1} \sum_q \epsilon^q
\Threej{j'}{1}{j}{-m'}{q}{m}.
\end{equation}
\end{widetext}

\noindent
Expressions in terms of intensity or laser power/waist may be worked out
with the aid of Eq. (\ref{expt_Rabi}).

The case in which the excited state can decay to multiple $n$ or $j$ levels is
more complicated, and the reader is referred to Ref. \cite{Weissbluth} or Ref
. \cite{Sobelman} for more information.  However, we \emph{will} deal with the
case of multiple ground-state hyperfine levels in Sec. \ref{HFS}.

\subsection{Breaking down to orbital angular momentum states}
\label{ls}

For theorists, there is still work to be done in relating Eq. (\ref{3jRabi}) to
theoretical calculations of atomic wave functions.  Eq. (\ref{3jRabi})
expresses the Rabi frequency in terms of the reduced matrix element $\langle
n'j'm'||\hat{r}\mathbf{\mathsf{C}^{(1)}}||njm\rangle$.  However, the
$\hat{r}\mathsf{C}^{(1)}$ only affects the \emph{spatial} part of the electron
state and leaves the spin alone.  Thus, in calculating Rabi frequencies
from scratch, we would like to break down the angular momentum into its
constituent parts:  $\mathbf{\hat{J}}=\mathbf{\hat{L}}+\mathbf{\hat{S}}$.  We
can re-express $\hat{r}\mathsf{C}^{(1)}$ more accurately as the tensor
product of $\hat{r}\mathsf{C}^{(1)}$ and the identity operator $\mathbb{I}_s$
acting on the spin state.  So we are interested in calculating

\begin{equation}
 \label{breaking}
	\bra{n'l's'j'}|\hat{r}\mathsf{C}^{(1)}\otimes\mathbb{I}_s|\ket{nlsj}.
\end{equation}

We can simplify this calculation by using Eq. (C.89) of Messiah.\cite{Messiah} 
In the present notation:

\begin{align}
 \label{decoupling}
	\bra{n'l's'j'}|\hat{r}\mathsf{C}^{(1)}\otimes\mathbb{I}_s|\ket{nlsj} = 
	\delta_{s,s'}\bra{n'l'}|\hat{r}\mathsf{C}^{(1)}|\ket{nl}
\nonumber\\
	\times (-1)^{j+l'+s'+1}\sqrt{(2j'+1)(2j+1)}\Sixj{l'}{1}{l}{j}{s'}{j'}.
\end{align}

\noindent
Now, $\hat{r}$ acts only on the radial part of the wave function, and
$\mathsf{C}^{(1)}$ acts only on the angular part.  So
$\bra{n'l'}|\hat{r}\mathsf{C}^{(1)}|\ket{nl}=\bra{n'}\hat{r}\ket{n}\bra{l'}
|\mathsf{C}^ {(1)}|
\ket{l} = \mathcal{R}_{n'l'}^{nl}\bra{l'}|\mathsf{C}^{(1)}|\ket{l}$.  Here
$\mathcal{R}_{n'l'}^{nl}$ is the radial integral
$\int R_{n'l'}^*(r)rR_{nl}\;r^2dr$, where the radial wave function $R_{nl}(r)$
is the output of the theoretical calculation of the electronic wave function.

It remains to evaluate $\bra{l'}|\mathsf{C}^{(1)}|\ket{l}$.  To do this, note
that, by the Wigner-Eckart theorem,

\begin{equation}
	\bra{l',0}C_0^1\ket{l,0}=(-1)^{l'}\bra{l'}|\mathsf{C}^{(1)}|\ket{l}
	\threej{l'}{1}{l}{0}{0}{0}.
\end{equation}

\noindent
On the other hand, using Eq. (C.16) of Messiah:\cite{Messiah}

\begin{align}
\bra{l',0}C_0^1\ket{l,0}&=\bra{l',0}\sqrt{\frac{4\pi}{3}}Y_0^1\ket{l,0}
\nonumber\\
		& = (-1)^0\sqrt{\frac{4\pi}{3}}\int Y_{0}^{l'} Y_0^1
Y_{0}^{l}\;d\Omega\nonumber\\
		& = \sqrt{(2l'+1)(2l+1)}\threej{l'}{1}{l}{0}{0}{0}
		\threej{l'}{1}{l}{0}{0}{0}.
\end{align}

\noindent
Comparing these expressions, we see that:

\begin{equation}
 \label{l_reduced}
\bra{l'}|\mathsf{C}^{(1)}|\ket{l} =
(-1)^{-l'}\sqrt{(2l'+1)(2l+1)}\threej{l'}{1}{l}{0}{0}{0},
\end{equation}

\noindent
At this point, it may be worth working out the explicit value of the 3-j
symbol.  From Table 2 of Edmonds\cite{Edmonds_L},\footnote{\label{Sobelman_3l}
Note that this expression disagrees in sign with Eq. (4.136) of
Sobelman.\cite{Sobelman}  However, Sobelman's equation may be a misprint, as it
disagrees Eq. (4.55) of his own book, which equation is consistent with Edmond's
Table 2.\cite{Edmonds_L}  The inconsistency leads to an incorrect sign in
Sobelman's Eq. (4.138), where the phase \emph{should} be $(-1)^{l_>}$.}, with
the projection numbers set to $0$\,

\begin{equation}
 \threej{l+1}{1}{l}{0}{0}{0}=(-1)^{l-1}\sqrt{\frac{(l+1)}{(2l+3)(2l+1)}}.
\end{equation}

\noindent
Now, $l\rightarrow l\pm 1$ in our transition, which means that,
if we use the symbol $l_>$ to denote the larger of $l'$ and $l$,

\begin{equation}
\label{0_3l}
 \threej{l'}{1}{l}{0}{0}{0} = (-1)^{l_>}\sqrt{\frac{l_>}{(2l'+1)(2l+1)}}.
\end{equation}

\noindent
This, in turn, implies that

\begin{equation}
 \label{simple_l_reduced}
	\bra{l'}|\mathsf{C}^{(1)}|\ket{l} = (-1)^{l_>-l'}\sqrt{l_>}.
\end{equation}

Finally, (dropping the $\delta_{s,s'}$ with the understanding that it is
implicit)

\begin{align}
\label{J_RME}
 	\bra{n'l's'j'}|\hat{r}\mathsf{C}^{(1)}\otimes\mathbb{I}_s|\ket{nlsj} = 
	(-1)^{j+l_>+s'+1}\mathcal{R}_{n'l'}^{nl}\nonumber\\
	\times\sqrt{(l_>)(2j'+1)(2j+1)} \sixj{l'}{1}{l}{j}{s'}{j'}.
\end{align}

\noindent
The interpretation of the 6-j symbol was discussed when these symbols were
first introduced in Sec. \ref{coupling}.  The 3-j symbol is present
simply because we must express the reduced-matrix element via the
Wigner-Eckart theorem in terms of \emph{some} (non-reduced) matrix
element, and we chose above to represent it in terms of
$\bra{l',0}C_0^1\ket{l,0}$.  The various square roots arise from the
normalization of the 3-j and 6-j symbols.

Finally, we can put the above together with Eq. (\ref{3jRabi}) for the complete
but somewhat lengthy expression:

\begin{align}
\label{decoupled_Rabi}
 	&\Omega_{e\leftarrow g}
=(-1)^{j'+j+l_>+s'+1-m'}\mathcal{R}_{n'l'}^{nl}\frac{eE_0}{\hbar}\nonumber\\
	&\times\sqrt{(l_>)(2j'+1)(2j+1)}\
	\sixj{l'}{1}{l}{j}{s'}{j'}\nonumber\\
	&\times	\sum_q \epsilon^q\Threej{j'}{1}{j}{-m'}{q}{m}.
\end{align}

\noindent
The quantity $E_0$ is given to us by the experimentalist, as is the
relative orientation of the laser polarization and the quantization axis
(typically due to the applied ``quantization'' magnetic field).  The
quantity $\mathcal{R}_{n'l'}^{nl}$ is given to us by the theorist.  The rest of
the quantities are specified purely by the geometry and are independent of the
details of the system.

\subsection{Rabi frequencies in the case of hyperfine structure}
\label{HFS}

The case of an atom with hyperfine structure (due to nuclear angular momentum
$\mathbf{\hat{I}}$) is somewhat more complicated than the above cases.  However,
the idea is the same. The Wigner-Eckart theorem still holds, and so Eq.
(\ref{3jRabi}) still applies, but with $j$ replaced with the total angular
momentum quantum number $F$ (where
$\mathbf{\hat{F}}=\mathbf{\hat{I}}+\mathbf{\hat{J}}$).  Thus:

\begin{align}
 \label{basic_hyperfine_Rabi}
 	\Omega_{e\leftarrow g} 
		 &=\frac{eE_0}{\hbar}(-1)^{F'-m_F'}\langle n'F'||
\hat{r}\mathbf{\mathsf{C}^{(1)}}||nF\rangle\nonumber\\
	 &\times\sum_q \epsilon^q \Threej{F'}{1}{F}{-m_F'}{q}{m_F}.
\end{align}

As before, the laser (in the electric dipole approximation) interacts only with
the orbital-angular-momentum part $\mathbf{\hat{L}}$ of the electronic
angular momentum $\mathbf{\hat{J}}=\mathbf{\hat{L}}+\mathbf{\hat{S}}$.  Writing
the quantum number $I$ first in labelling the states, we use Eq. (C.90) of
Messiah,\cite{Messiah} to write:

\begin{align}
\label{hyperfine_RME}
 	&\langle n'F'||\hat{r}\mathbf{\mathsf{C}^{(1)}}||nF\rangle
\equiv
	\langle n'I'j'F'||\hat{r}\mathbf{\mathsf{C}^{(1)}}||nIjF\rangle
\nonumber\\
&=\delta_{I',I}\delta_{j',j}(-1)^{F'+I'+j+1}\sqrt{(2F'+1)(2F+1)}\nonumber\\
	&\times\sixj{j'}{1}{j}{F}{I'}{F'}\langle n'j'||
\hat{r}\mathbf{\mathsf{C}^{(1)}}||nj\rangle.
\end{align}

\noindent
The 6-j symbol, roughly speaking, accounts for the probability (amplitude) that
one can change the overall angular momentum from $F$ to $F'$ by the $1$ unit
 of photon angular momentum by changing the \emph{electron's} angular
momentum from $j$ to $j'$ (since the laser field only couples to the electron).

Using Eq. (\ref{J_RME}) to express the value of 
$\langle nj||\hat{r}\mathbf{\mathsf{C}^{(1)}}||n'j'\rangle$, it follows that the
Rabi frequency in the case of hyperfine structure is:

\begin{widetext}
\begin{align}
 \label{hyperfine_Rabi}
 	\Omega_{e\leftarrow g}&=\frac{eE_0}{\hbar}(-1)^{2F'+I'+j+1-m_F'}
	\sqrt{(2F'+1)(2F+1)} \sixj{j'}{1}{j}{F}{I'}{F'}
	\langle n'j'||\hat{r}\mathbf{\mathsf{C}^{(1)}}||nj\rangle
	 \sum_q \epsilon^q \Threej{F'}{1}{F}{-m_F'}{q}{m_F}\nonumber\\
	&=\frac{eE_0}{\hbar}(-1)^{2F'+I'+2j+l_>+s'-m_F'}\mathcal{R}_{n'l'}^{nl}
	\sqrt{(l_>)(2j'+1)(2j+1)(2F'+1)(2F+1) }	\nonumber\\
	&\times\sixj{l'}{1}{l}{j}{s'}{j'}\sixj{j'}{1}{j}{F}{I'}{F'}
	 \sum_q \epsilon^q \Threej{F'}{1}{F}{-m_F'}{q}{m_F}.
\end{align}
\end{widetext}
\noindent
(The delta functions have been supressed for the sake of brevity, and I've
used the fact that $(-1)^2=1$.)

\section{Conclusion}
\label{conclusion}

In the end, then, the interaction of an atom with an applied laser field
induces a dipole moment of magnitude $e\mathcal{R}_{n'l'}^{nl}$ in the atom. 
The interaction between the dipole moment and the field then drives transitions
between different atomic levels.  Since the interaction has well-defined
rotational symmetry, angular momentum must be conserved overall.  The
transition probability is thus ``modulated'' by the probability amplitude for
angular momentum to be conserved in a particular transition, depending on the
relative orientation of the atom and the applied field.  This ``modulation'' is
embodied by the Wigner-Eckart theorem which, if you will, ``splits up'' the
transition probability amongst the different states whose coupling conserves
angular momentum.  The total transition rate out of an excited state (driven by
vacuum fluctuations) is the same for all states in a given degenerate angular
momentum manifold, as it must be.  This latter rate is given by the Einstein
$A$ coefficient, and allows connection with experimentally determined
quantities.

\appendix
\section{Comparison with other work}
\label{comparison}

Some or all of the results in this paper may be found scattered throughout the
literature.  However, it can be challenging to compare results found in
different works.  This is due in part to different systems of units,
different choices of active vs. passive rotations, different definitions of
reduced matrix elements in the Wigner-Eckart theorem, or different
arrangements of the elements of 3-j and 6-j symbols, but most of all to
differences in sign/phase conventions.  As long as the reader picks one
convention \emph{and sticks with it}, results will be self-consistent - barring
algebra errors along the way!  When algebra errors occur, it is inevitably 
in determining the sign of the matrix elements.

Two other works which succinctly express Rabi frequencies in the case of fine
and/or hyperfine structure are Metcalf and van der Straten\cite{CoolingNeutrals}
and Farell and MacGillivray.\cite{Farell_MacGillivray}

Metcalf and van der Straten's Eq. (4.32) is consistent with Eqs. (\ref{3jRabi})
and (\ref{decoupling}) of this work.  However, there are sign issues with Eqs.
(4.26) and (4.27) of Metcalf and van der Straten.  In the first equation, their
$\sqrt{\frac{4\pi}{3}}\int\sin\theta d\theta d\phi Y_{l'm'}(\theta,\;\phi)
Y_{1q}(\theta,\;\phi)Y_{lm}(\theta,\;\phi)$ \emph{should} be
$\sqrt{\frac{4\pi}{3}}\int\sin\theta d\theta d\phi
Y_{l'm'}^*(\theta,\;\phi)Y_{1q}(\theta,\;\phi)Y_{lm}(\theta,\;\phi)$ (note the
complex conjugation).  Furthermore, their Eq. (4.27) does not always agree in
sign with the properly expressed integral of the three spherical harmonics.  In
addition, their expression (4.33) is not consistent in sign with Eq.
(\ref{hyperfine_Rabi}) of this work.

Farrell and MacGillivray write their states as $\ket{nslj}$ rather than
$\ket{nlsj}$ as is done in the present work.  This produces a differerent
overall sign.  However, if the reader consistently applied their convention,
self-consistent results would ensue - \emph{except} that Farrell and
MacGillivray use Eq. (4.136) of Sobelman, which is incorrect as pointed out in
Sec. \ref{ls}. Thus, though their results will be self-consistent for
transistions between fixed $l$, $l'$ they could be inconsistent if used to treat
simultaneous coherently driven excitations to multiple $l'$ levels.

Eq. (\ref{decoupling}) for the
$\bra{n'l's'j'}|\hat{r}\mathbf{\mathsf{C}^{(1)}}|\ket{nlsj}$ agrees in magnitude
and sign with Eq. (23.1.24) of Weissbluth\cite{Weissbluth} and with Eq. (14.54)
of Cowan,\cite{Cowan_book} and in agrees in magnitude with Eq. (9.63) of
Sobelman\cite{Sobelman} (who uses $\ket{nslj}$ rather than $\ket{nlsj}$).  Eq.
(\ref{l_reduced}) for the reduced matrix element
$\bra{l'}|\mathbf{\mathsf{C}^{(1)}}|\ket{l}$ agrees with Eq. (14.55) of Cowan
but, as discussed previously, disagrees with Eq. (4.126) of Sobelman.

Issues of different or even inconsistent minus signs become irrelevent when one
calculates incoherent rates.  Thus, for example, Eq. (\ref{J_lifetime})
for the Einstein A coefficient agrees with Eq. (14.32) of
Cowan,\cite{Cowan_book} and Eq. (9.47) of Sobelman\cite{Sobelman} (though Cowan
expresses his result in terms of wavenumbers, and Sobelman uses CGS units).

\section{Sign of Eq. (\ref{RME_phase})}
\label{proof}

From Eq. (\ref{J_RME}), the phase of
$\bra{n'j'}|\hat{r}\mathbf{\mathsf{C}^{(1)}}|\ket{nj}$ is determined by the
sign of $(-1)^{j+s'+1}\sixj{l'}{1}{l}{j}{s'}{j'}\threej{l'}{1}{l}{0}{0}{0}$. 
It is possible to evaluate this phase by employing cautious reasoning and the
fact that in an electric-dipole transition, $l\rightarrow l\pm1$ and that
$j\rightarrow j\pm1,\;0$.  Note, however, that transitions in which
$j\rightarrow j\pm 1$ but $l\rightarrow l\mp 1$ do not occur - such transitions
do not satisfy the triangle relations necessary for the 6-j symbol to be
non-zero.\cite{Messiah}

We can determine the sign of the 6-j symbol on a case-by-case basis using the
symmetry properties of the 6-j symbols and Table 5 of
Edmonds\cite{Edmonds_L}(which is also available in other forms in other
references).  First, note that, permuting the columns of the
six-j symbol, and then flipping the rows of the resulting first and second
columns, $\sixj{l'}{1}{l}{j}{s'}{j'}=\sixj{1}{l}{l'}{s'}{j'}{j}
=\sixj{s'}{j'}{l'}{1}{l}{j} .$  This form is suitable for comparison with
Edmonds.

For the case $j'=j+1$, $l'=l+1$ we have that $j'>j$, $l'>l$ and

\begin{equation}
 \sixj{s'}{j'}{l'}{1}{l}{j}=\sixj{s'}{j'}{l'}{1}{l'-1}{j'-1}\propto
(-1)^{j'+l'+s'}=(-1)^{j_> + l_> + s'}
\end{equation}

\noindent
For the case $j'=j$, $l'=l+1$ we have that $l'>l$ and

\begin{equation}
\sixj{s'}{j'}{l'}{1}{l}{j}=\sixj{s'}{j'}{l'}{1}{l'-1}{j'}\propto
(-1)^{j'+l'+s'}=(-1)^{j_> + l_> + s'}.
\end{equation}

\noindent
For the case $j'=j$, $l'=l-1$, we have that $l>l'$ and

\begin{equation}
 \sixj{s'}{j'}{l'}{1}{l}{j}=\sixj{s'}{j}{l}{1}{l-1}{j}
\propto (-1)^{j+l+s'}=(-1)^{j_> + l_> + s'}.
\end{equation}

\noindent
And finally, for the case $j'=j-1$, $l'=l-1$, we have that $j>j'$, $l>l'$ and

\begin{equation}
\sixj{s'}{j'}{l'}{1}{l}{j}=\sixj{s'}{l}{j}{1}{j-1}{l-1}
\propto (-1)^{j+l+s}=(-1)^{j_> + l_> + s'}.
\end{equation}

\noindent
So in all cases,
$\sixj{s'}{j'}{l'}{1}{l}{j}\propto=(-1)^{j_> + l_> + s'}$.

As for the 3-j symbol, Table 2 of Edmonds\cite{Edmonds_L} indicates that
$\threej{l'}{1}{l}{0}{0}{0}\propto(-1)^{(l'+l+1)/2}$.  Now, given that $l'=l\pm
1$, $(l'+l+1)/2=[(l\pm1)+l+1]/2$ which is either $(2l+2)/2$ (if $l'=l+1$) or
$2l/2$ (if $l'=l-1$).  So in either case,
$\threej{l'}{1}{l}{0}{0}{0}\propto(-1)^{l_>}$.

Putting these results together, we have that
$\bra{nj}|\hat{r}\mathbf{\mathsf{C}^{(1)}}|\ket{n'j'}\propto
(-1)^{j+2s'+1+j_>+2l_>}$.  Since $s'=1/2$ and $l_>$ is an integer, $(-1)^{2s'+1
+2l_>}=1$.  So finally, we have that
$\bra{n'j'}|\hat{r}\mathbf{\mathsf{C}^{(1)}}|\ket{nj}\propto
(-1)^{j+j_>}$, as assumed in Eq. (\ref{J_RME}).

%
%
%
%
%

\acknowledgements{I thank Jason Nguyen, Laura Toppozini, Duncan O'Dell, A.
Kumarakrishnan and particularly Ralph Shiell, for helpful discussions and/or
critical readings of the manuscript and Malcolm Boshier for unpublished notes
which clearly laid out some steps left murky in other works.  This work was
supported by NSERC.}

\bibliography{/home/kingb/research/papers/Brian_Mac}

\end{document}